\documentclass[12pt]{article}

\title{Tempo and mode in quasispecies evolution}

\author{Joachim Krug$^{1,2,3,4}$ \\
{\small 
1. Fachbereich Physik, Universit\"at Essen, D-45117 Essen, Germany\footnote{Permanent address.}} \\
{\small 2. CAMP and Department of Physics, DTU, DK-2800 Kongens Lyngby, Denmark} \\
{\small 3. Niels Bohr Institute, Blegdamsvej 17, DK-2100 Copenhagen \O, Denmark} \\
{\small 4. Institute for Theoretical Physics, UCSB, Santa Barbara, 
CA 93106-4030, USA}
}

\begin{document}

\maketitle             

\begin{abstract}
Evolutionary dynamics in an uncorrelated rugged fitness landscape is studied
in the framework of Eigen's molecular quasispecies model. We consider
the case of strong selection, which is analogous to
the zero temperature limit
in the equivalent problem of directed polymers in random
media. In this limit the population is always localized at a single
temporary master sequence $\sigma^\ast(t)$, and we study the statistical
properties of the evolutionary trajectory which $\sigma^\ast(t)$ traces
out in sequence space.
Numerical results for binary sequences
of length $N=10$ and exponential and uniform
fitness distributions are presented.
Evolution proceeds by intermittent jumps between
local fitness maxima, where high lying maxima are visited more frequently
by the trajectories. The probability distribution for the total time
$T$ required to reach the global maximum shows a  $T^{-2}$-tail, which
is argued to be universal and to derive from near-degenerate
fitness maxima. The total number of jumps along any given trajectory
is always small, much smaller than predicted by the statistics
of records for random long-ranged evolutionary jumps.

\end{abstract}

\newpage

\small

\noindent
{\it ``The concept of quasispecies is not just a model that involves
any odd assumption; it shows how to view the darwinian world of replicating
and mutating species from a physical viewpoint.''}
\hfill M. Eigen \cite{eigen96}

\normalsize

\section{Introduction and motivation}

Eigen's quasispecies theory of molecular evolution
is the simplest mathematical model that incorporates the
central Darwinian paradigm of natural selection acting on
variability created by random mutations.
The model was originally developed to understand the conditions for the 
maintenance of information in systems of self-instructive
replicating macromolecules \cite{eigen71}. Such systems can be realized
in the laboratory in the form of populations of RNA strands
which replicate {\em in vitro} in the presence of RNA replicase,
displaying a wide range of evolutionary phenomena \cite{spiegelman,biebricher83,biebricher97,mccaskill93}.
The notion of a {\em quasispecies} \cite{eigen89} refers to 
the structure of self-replicating populations, which 
typically consist of a distribution of related mutants centered around a
most abundant {\em master} genotype (see below). The quasispecies
structure plays an important role in the evolution of RNA viruses, 
where the presence of a wide range of mutants
allows the virus to adapt rapidly to environmental changes    
\cite{domingo98,sole99}. On the other hand, the existence of
an error threshold beyond which no localized quasispecies can be 
maintained (see Eq.(\ref{error})) places an upper bound on the
genome length of RNA viruses \cite{eigen88}. 

The mathematical structure of the quasispecies model  
has made it a favored entrance
way for statistical physicists into the field of biological 
evolution\footnote{Recent articles which review this connection
are \cite{peliti97,baake99,drossel01}.}. It was first
observed by Leuth\"ausser \cite{leuthausser87} that the
discrete time dynamics (\ref{quasi}) can be interpreted as 
a transfer matrix of a two-dimensional Ising model, where
the genotype sequences become one-dimensional spin configurations
that are coupled in the time direction through the mutation matrix
(\ref{Q}) \cite{tarazona92}. A similar relation can be established
between (\ref{quasi}) and the transfer matrix of a polymer directed
along the time axis \cite{galluccio96,galluccio97}. In addition, Baake
and coworkers have recently exploited the equivalence between
quantum spin chains and a class of kinetic evolution equations closely
related to the quasispecies model, in which mutation and selection
occur in parallel \cite{baake99,baake97}.  

In its most basic version, the quasispecies model
is formulated in terms of standard
chemical reaction equations\footnote{Similar equations
arise in classical population genetics \cite{baake99}.} 
written for the concentrations
$n_\sigma(t)$ of sequences $\sigma = (\sigma_1,...,\sigma_N)$,
each of which is composed of $N$
symbols drawn from an alphabet of $K$ letters; the usual choice is
a binary alphabet ($K=2$), so that
$\sigma_i$ takes the values 0 and 1. 
The resulting {\em sequence space} consists then of 
$S = 2^N$ points arranged on the vertices of an $N$-dimensional hypercube. 
Each sequence $\sigma$
reproduces at a rate $W(\sigma)$, which may be taken as a measure of
its fitness \cite{peliti97}. 
In the reproduction process errors occur with a mutation
probability $\mu$ per site. The probability of creating a sequence
$\sigma'$ when attempting to copy sequence $\sigma$ is therefore equal
to
\begin{equation}
\label{Q}
Q_\mu(\sigma \to \sigma') = \mu^{d_H(\sigma, \sigma')} (1 - \mu)^{N -
d_H(\sigma, \sigma')}
\end{equation}
where
\begin{equation}
\label{hamming}
d_H(\sigma, \sigma') = \sum_{i=1}^N (\sigma_i - \sigma_i')^2
\end{equation}
is the Hamming distance between the two sequences,
i.e. the number of digits in which the two differ. 
The dynamical evolution
in discrete time is then given by
\begin{equation}
\label{quasi}
n_\sigma(t+1) = \sum_{\sigma'} W(\sigma') Q_\mu(\sigma' \to \sigma)
n_{\sigma'}(t).
\end{equation}
Introducing the constraint of a fixed number of molecules leads to 
nonlinear loss terms on the right hand side of (\ref{quasi}) 
\cite{eigen89}.
However since these can generally be transformed away, we ignore
this complication here, at the expense of dealing with exponentially
growing population numbers.  

The linear form of the evolution equation (\ref{quasi}) makes it plain
that, for long times, the concentrations will approach that eigenvector
of the evolution matrix $W(\sigma) Q_\mu(\sigma \to \sigma')$
which corresponds to the largest eigenvalue. Provided this eigenvector
is {\it localized} in sequence space, it
defines the quasispecies: A distribution of related mutants
centered around the master sequence, which usually is the
sequence with the maximum replication rate $W(\sigma)$.
The most celebrated property of the model is its prediction
of a sharp {\it error threshold}, where the quasispecies delocalizes,
and the population spreads uniformly over sequence space. 
In terms of the sequence length $N$ and
the mutation probability $\mu$, the condition for a localized
quasispecies takes the form \cite{eigen71,eigen89}
\begin{equation}
\label{error}
N < N_{\rm max} = \frac{\ln A}{\mu},
\end{equation}
where $A$ denotes the {\it selective advantage}, a measure
for the superiority of the master sequence compared to the other
sequences. For the simplest case of a single peak fitness landscape,
where the master sequence replicates at rate $W_0$ and all other
sequences replicate at rate $W_1 < W_0$, the selective advantage
is $A = W_0/W_1$, while for randomly distributed replication
rates it is a functional of the rate distribution \cite{mccaskill84a,franz93}.
In terms of the physical analogies described above, the
error threshold phenomenon is equivalent to the
thermal phase transition in the Ising model
\cite{leuthausser87,tarazona92,baake97,franz93,franz97}
and to the thermal unbinding
of a directed polymer bound to an attractive columnar defect
along the time direction 
\cite{galluccio96,galluccio97}.

Much less appears to be known about the evolutionary dynamics of
the model, that is, the approach to the final quasispecies distribution
from an initial localized or delocalized state. It was first pointed
out by McCaskill \cite{mccaskill84a,mccaskill84b} that this 
dynamics should take the form of a ``slowing optimization walk''
through a succession of metastable states which correspond, in some
sense, to local maxima in the fitness landscape. The separation of
time scales between the (long)
residence time in a metastable state
and the (brief) transition time to the next maximum implies a 
{\em punctuated} pattern of evolution \cite{simpson,newman85,gould93},
which can be analyzed in analogy to variable range hopping in
condensed matter physics \cite{ebeling84,jkhh93,zhang97}. 

We should concede from the outset that the deterministic rate
equations (\ref{quasi}) are not an entirely appropriate 
description for this evolutionary regime, since the transition
between two local maxima proceeds through the tails
of the localized, metastable quasispecies, where the population
numbers are small and fluctuations due to the finite number of 
molecules cannot be ignored \cite{mccaskill84b,zhang97,alves98}.
It seemed nevertheless worthwhile to explore these questions
within the most basic, deterministic model, before turning
to more sophisticated approaches. 

The present paper reports on some preliminary results from such
an investigation. To avoid the complications due to a finite
error threshold, we consider a strong selection limit (to be 
described in the next section), in which the population is localized
at a single site in sequence space at all times. This allows
a direct comparison with simple schematic models of evolutionary
dynamics, such as adaptive walks 
\cite{kauffman87,macken89,macken91,flyvbjerg92} and record
dynamics \cite{sibani95,sibani97,sibani98}.
Adaptive walks describe the evolution of a genetically homogeneous
population under the assumption that deleterious mutations 
(which decrease the fitness) are
eliminated, while advantageous mutations spread 
instantaneously. The population then performs an uphill walk
in the fitness landscape, which terminates at a local maximum
where no fitter one-step neighbors are available. 
In contrast, in the quasispecies model
the population evolves through a chain
of local fitness maxima 
which progresses all the way to  
the {\it global} optimum. 
Some qualitative properties of these 
{\it evolutionary trajectories} are described
in Section \ref{Trajectories}, while Section \ref{Times}
focuses on a specific statistical feature, the total evolution time.
A comparison with record dynamics is provided in Section \ref{Records},
and some open questions are formulated in Section \ref{Outlook}.

\begin{figure}[htb]
\centering
\vspace*{75mm}
\includegraphics{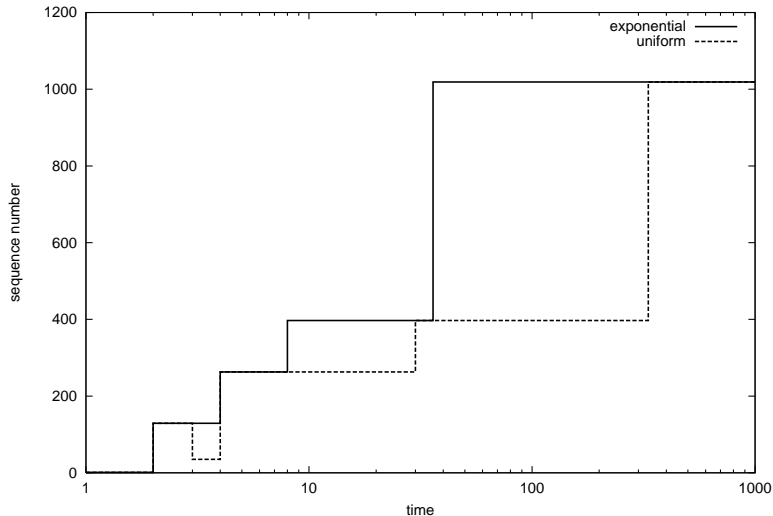}
\caption[]{Two evolutionary trajectories generated in two
fitness landscapes with identical ordering of fitnesses but
different fitness distributions. The monotonic increase
in sequence number for the exponential distribution is
fortuitous - there is no correspondence between the position
of a sequence on the $y$-axis and its fitness}
\label{fig_traj}
\end{figure}

Throughout the paper we will consider maximally rugged fitness landscapes,
in which the reproduction rate $W(\sigma)$ is chosen independently and
randomly for each sequence. In contrast to simpler permutation 
invariant landscapes \cite{baake99}, this makes it necessary to 
store all $2^N$ sequences during the iteration of the evolution 
equations\footnote{An approximate scheme which reduces
the storage requirement from $2^N$ to $N$ is described
in \cite{karl}.}, 
restricting our numerical treatment to rather short sequences;
the results shown here are for $N=10$. A systematic analysis
of the dependence on sequence length will be presented elsewhere
\cite{karl}.          

\section{The strong selection limit}

The form of the strong selection limit is motivated by
the analogy with the zero temperature limit in the associated
problems of statistical physics.
Following Peliti \cite{peliti97,franz97} we introduce an inverse ``selective
temperature'' $k> 0$ by writing the reproduction rates in the form
\begin{equation}
\label{temp}
W(\sigma) = e^{k F(\sigma)}.
\end{equation}
We want to take the strong selection limit $k \to \infty$ in such a way
that only a few mutations occur in each time step. This requires to
scale the mutation rate as
\begin{equation}
\label{muscale}
\mu = e^{-k \gamma}
\end{equation}
where $\gamma > 0$ is a constant. Inserting (\ref{temp}) and (\ref{muscale})
into (\ref{quasi}) it is clear that the sequence concentrations will grow
for large $k$ as
\begin{equation}
\label{nk}
n_\sigma(t) = e^{k E(\sigma, t)}.
\end{equation}
In the limit $k \to \infty$ the evolution equation (\ref{quasi}) then reduces
to the recursion
\begin{equation}
\label{nonlocal}
E(\sigma, t+1) = \max_{\sigma'} [ E(\sigma', t) + F(\sigma') -
\gamma d_H(\sigma, \sigma')].
\end{equation}
Since the term $- \gamma d_H$ suppresses mutations to far away
sequences, we expect similar behavior for a model
in which only nearest neighbor mutations are allowed,
\begin{equation}
\label{local}
E(\sigma, t+1) = \max_{d_H(\sigma,\sigma') \leq 1}
[ E(\sigma', t) + F(\sigma') -
\gamma d_H(\sigma, \sigma')].
\end{equation}
All results shown in this paper were obtained using the
nearest neighbor rule (\ref{local}), with the parameter
$\gamma$ set to unity\footnote{For a discussion of the differences
between the rules (\ref{nonlocal}) and (\ref{local}),
see \cite{karl}.}.

We still need to specify the probability distribution $p(F)$ of the 
fitnesses $F(\sigma)$. Two choices will be considered: The exponential
distribution 
$p(F) = e^{-F}$, $F \geq 0$,
and a uniform distribution on the interval $[0,S]$, where
$S = 2^N$. The reason for
this particular scaling of the width of the uniform distribution will
become clear below in Section \ref{Times}.

\section{Evolutionary trajectories}
\label{Trajectories}

It is evident from (\ref{nk}) that, in the strong selection limit
$k \to \infty$, the entire population resides at the global maximum
of the function $E(\sigma,t)$. The position of this maximum in 
sequence space will be referred to as the master sequence at time
$t$, and denoted by $\sigma^\ast(t)$. At time $t=0$ the master sequence
is placed at a randomly chosen point $\sigma^{(i)}$ by
setting $E(\sigma^{(i)},t) = 0$ and $E(\sigma,t) = - \infty$ for
$\sigma \neq \sigma^{(i)}$. The subsequent time evolution $\sigma^\ast(t)$
defines an evolutionary trajectory.

Inspection shows that,
after one or two time steps, such a trajectory passes exclusively
through local fitness maxima, and eventually, after a total evolution 
time $T$, it invariably reaches the global fitness maximum.
During the evolution, the master sequence spends increasingly long
time intervals at local maxima of increasing fitness, with a few
abrupt transitions in between (Figure \ref{fig_traj}). 
The number of transitions is small (see Figure \ref{fig_trans}),
much smaller than the number of local fitness maxima, which is
on average equal to $2^N/(N+1) \approx 93$ \cite{kauffman87}.
This implies that most local maxima are bypassed by a typical
trajectory. 

\begin{figure}[htb]
\centering
\vspace*{75mm}
\includegraphics{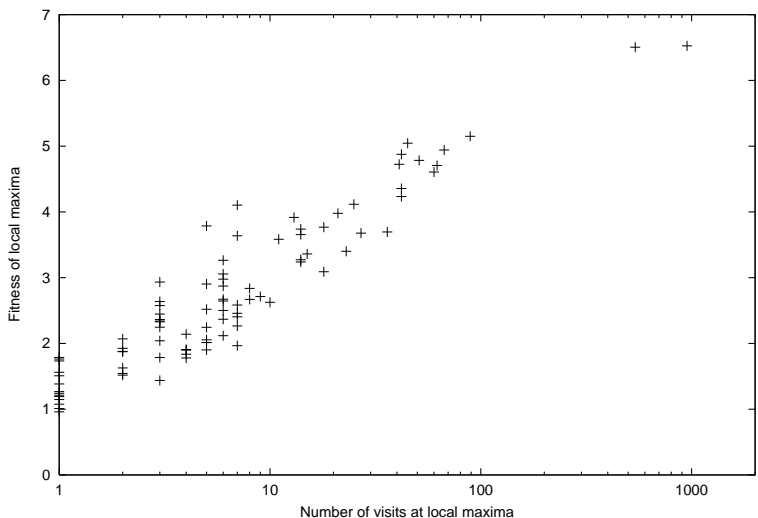}
\caption[]{Fitness $F(\sigma)$ vs. the number of visits for all
local maxima in a fitness landscape with exponential fitness
distribution $p(F)$. The particular landscape used here is near degenerate
(gap size $\epsilon \approx 0.02$)}
\label{fig_maxima}
\end{figure}

A quantification of this statement is shown in Figure
\ref{fig_maxima}, in which the fitness $F$ of local maxima
is plotted against the number of times it is visited by an
evolutionary trajectory. These data were generated by going through
all possible starting points $\sigma^{(i)}$ in a fixed fitness landscape.
The figure shows a roughly linear correlation between the fitness of a 
maximum and the logarithm of the number of visits.

In relation to adaptive walks 
\cite{kauffman87,macken89,macken91,flyvbjerg92}, which
respond only to the relative ordering of fitnesses and not to their
actual values, it is of interest to ask to what extent the set of
maxima visited by a given trajectory is determined by the ordering
of fitnesses. For this reason Figure \ref{fig_traj} shows
two trajectories evolving in landscapes which were generated 
using the same random numbers -- thus having identical ordering
of fitnesses -- but with different fitness distributions. 
It can be seen that the set of local maxima visited by the two
trajectories is almost identical, apart from a small detour taken
by the ``uniform'' trajectory, but the timing of the evolutionary
transitions is markedly different in the two cases. 
With reference to G.G. Simpson's classic treatise \cite{simpson},
we may say that the fitness distribution affects only the {\it tempo},
but not the {\it mode} of quasispecies evolution. 
For a quantitative analysis of the temporal aspects we next
turn to the distribution of evolution times.

\section{Distribution of evolution times}
\label{Times}

\begin{figure}[ht]
\centering
\vspace*{75mm}
\includegraphics{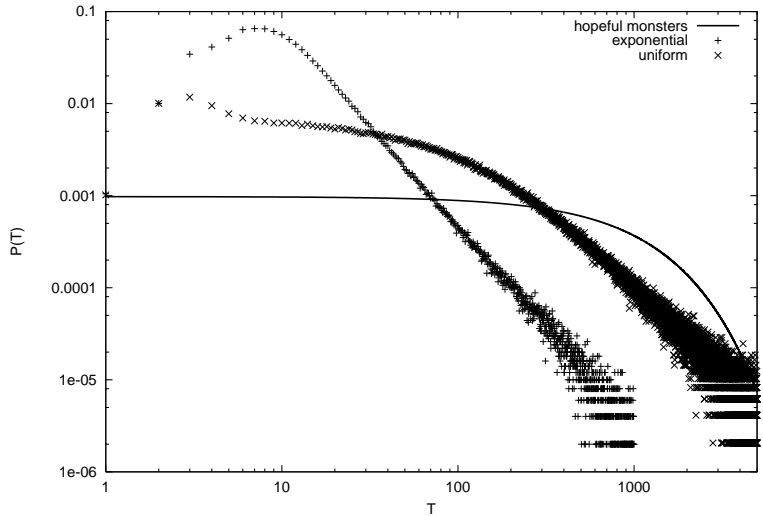}
\caption[]{Distribution of the total time $T$ required to reach the global
fitness maximum. Symbols show data obtained by averaging over 500000 uncorrelated
fitness landscapes with exponential (+) and uniform ($\times$) fitness distributions,
while the full line shows the distribution (\ref{PTrecords}) obtained for record
dynamics. The simulations with exponential fitness distribution
were stopped after 1000 time steps}
\label{fig_times}
\end{figure}

Figure \ref{fig_times} 
shows the distribution $P(T)$ of the number of time steps $T$
required to reach the global fitness maximum,
obtained by averaging over 500000 landscapes
with exponential and uniform fitness distributions.
The time distribution for the exponential case displays a distinct
maximum around $T = 7$, followed by a slowly decaying tail
which is well described by the power law
\begin{equation}
\label{power}
P(T) \sim T^{-2}
\end{equation}
over roughly two decades. The distribution for the uniform
case is much broader, but a similar power law tail can be
seen for times $T \geq 500$. 

The power law (\ref{power}) appears to be a simple consequence
of the order statistics of uncorrelated fitness landscapes.
Let $F^{(1)} > F^{(2)} > ...> F^{(S)}$ be a realization of fitnesses
arranged in decreasing order.
As a measure of the spread in fitnesses among the most fit
sequences we introduce the {\em fitness gap}
\begin{equation}
\label{eps}
\epsilon = F^{(1)} - F^{(2)} > 0,
\end{equation}
which is a random variable characteristic of each fitness landscape.
In the late stage of evolution the population will typically make a
a transition from the 
second best sequence $\sigma^{(2)}$ (or some local fitness
maximum with comparable fitness\footnote{In fact
$\sigma^{(2)}$ {\it is} a local
fitness maximum with high probability $1 - N/(2^N - 1) 
\approx 0.990$.}) to the
globally optimal sequence
$\sigma^{(1)}$. From the evolution rule (\ref{local})
it is easy to see that, for small $\epsilon$, this transition will require
a time of the order of 
\begin{equation}
\label{T}
T \approx a(N)/\epsilon,
\end{equation}
where the coefficient $a$ is determined by the early stages
of the evolution process \cite{karl}. Thus given the
gap distribution $P_{\rm g}(\epsilon)$
the tail of the distribution of evolution
times can be estimated to be
\begin{equation}
\label{Tdist}
P(T) \approx a T^{-2} P_{\rm g}(a/T),
\end{equation}
and a $T^{-2}$ power law follows for $T \gg a$, provided that
$0 < P_{\rm g}(0) < \infty$. The gap distribution is 
given by $P_{\rm g}(\epsilon) = 
e^{-\epsilon}$ {\em both} for exponentially distributed fitnesses,
and for fitnesses distributed uniformly between 0 and $S$, when
$S$ is large \cite{feller}. 
The striking difference between the two evolution
time distributions seen in Figure \ref{fig_times}
is related to the different scaling of the coefficient
$a(N)$ in (\ref{T}) with sequence length: In the exponential case
$a(N) \sim N^{3/2}$, while in the uniform case $a(N) \sim \sqrt{N} S$
\cite{karl}.

To compute $P_{\rm g}(0)$ for general fitness distributions,
note first that the joint distribution of
$F^{(1)}$ and $F^{(2)}$ is given by \cite{david}
\begin{equation}
\label{joint}
P_2(F^{(1)}, F^{(2)}) =  S(S-1) p_c(F^{(2)})^{S-2} p(F^{(1)}) p(F^{(2)})
\end{equation}
where $p_c(F) = \int_0^F dF' \; p(F')$ denotes the cumulative fitness
distribution. The cumulative gap distribution is obtained by integration,
$$ {\rm Prob}[F^{(1)} - F^{(2)} < \epsilon] = $$
\begin{equation}
\label{gapcum}
  S(S-1)
\int_0^\infty d F^{(2)} \; p(F^{(2)}) p_c(F^{(2)})^{S-2}
\int_{F^{(2)}}^{F^{(2)} + \epsilon} d F^{(1)} \; p(F^{(1)}),
\end{equation}
which tends to $P_{\rm g}(0) \epsilon$
for $\epsilon \to 0$. Thus we conclude that
\begin{equation}
\label{pg0}
P_g(0) =  S(S-1) \int_0^\infty dF p(F)^2 p_c(F)^{S-2}
\end{equation}
which is clearly finite and nonzero.

\begin{figure}[ht]
\centering
\vspace*{75mm}
\includegraphics{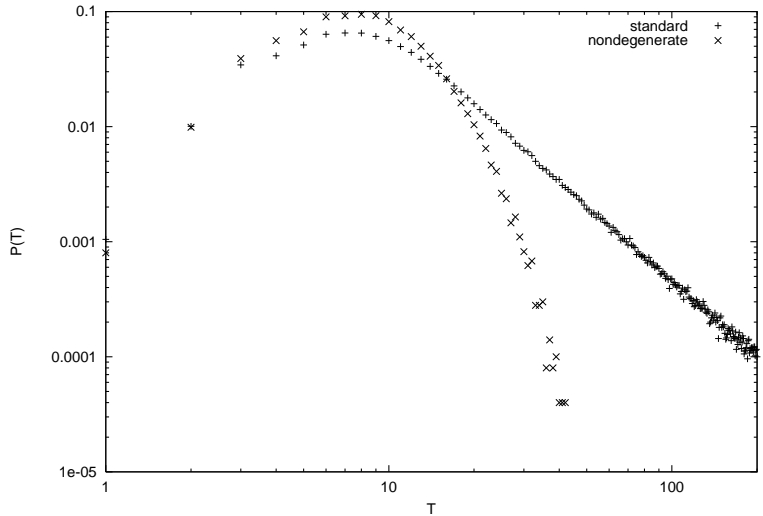}
\caption[]{Comparison of the distribution of evolution times
for exponentially distributed fitnesses (+) with an ensemble
of landscapes for which near-degeneracies (small gaps) have
been removed ($\times$). The latter data were 
averaged over 50000 realizations}
\label{fig_nondegen}
\end{figure}

The relationship (\ref{Tdist}) implies that the near-degenerate
fitness landscapes, which have very small gaps, are the ones 
that give rise to anomalously long evolution times. Figure
\ref{fig_nondegen} illustrates this connection. The data 
shown as crosses were obtained from an average over 
exponential fitness landscapes,
for which the fitness gap $\epsilon$ was increased artificially
by increasing the global fitness maximum according to
$F^{(1)} \to F^{(1)} + 1$. This is seen to immediately remove
the power law tail (\ref{power}).

\section{Comparison to record dynamics}
\label{Records}

A simple schematic analogue of the nonstationary (ever slowing) evolution process
found in the quasispecies model is provided by the dynamics of records
\cite{sibani95,sibani97,sibani98,feller}, which is equivalent to evolution by
long-ranged random mutations known in the classical literature as the
theory of ``hopeful monsters'' \cite{simpson,kauffman87}.
In the present context it reduces to the following rule for the motion
of the master sequence $\sigma^\ast(t)$ in sequence space: At each time step, the population
attempts a jump to another, randomly chosen sequence $\sigma' \neq \sigma^\ast(t)$.
The move is accepted, and $\sigma^\ast(t+1) = \sigma'$, if $F(\sigma') > F(\sigma^\ast(t))$; otherwise
it is discarded and $\sigma^\ast(t+1) = \sigma^\ast(t)$.
Thus the current sequence $\sigma^\ast(t)$ represents the {\it fitness record} among the
sequences which the population has encountered so far.

\begin{figure}[hbt]
\centering
\vspace*{75mm}
\includegraphics{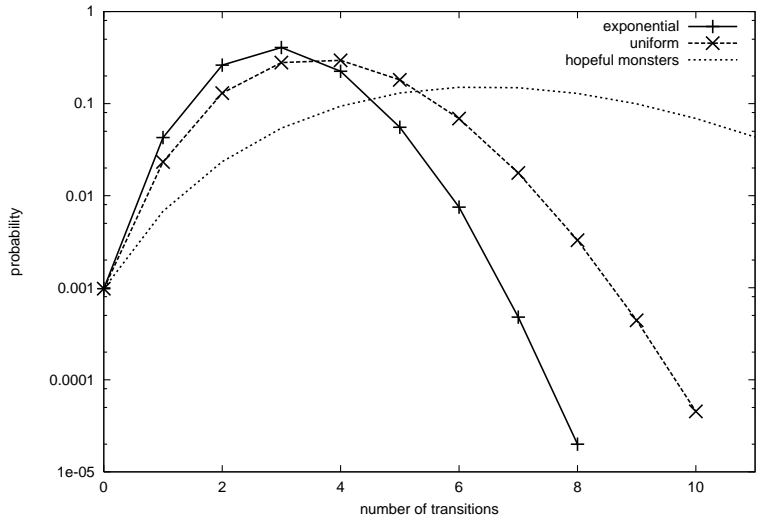}
\caption[]{Probability distributions for the total number of evolutionary jumps required to reach
the global fitness maximum. Symbols show data obtained by averaging over 500000 fitness landscapes
with exponential (+) and uniform ($\times$) fitness distributions, while the dotted
line is the log-Poisson distribution 
(\ref{Pnrecords}) predicted by record dynamics}
\label{fig_trans}
\end{figure}

Clearly this process gives rise to a
step-like, punctuated pattern which is {\it qualitatively} similar to that shown above in Figure
\ref{fig_traj}. Here we are concerned with a {\it quantitative} comparison of statistical
properties. Let us first compute the probability distribution of the total
evolution time $T$ for the record dynamics. Since the probability of finding the global fitness
maximum in any jump is $1/S$, the probability that it has not been found up to time $t$ is
$(1 - 1/S)^t \approx e^{-t/S}$ for large $S$ and $t$. Taking the derivative one obtains
\begin{equation}
\label{PTrecords}
P(T) = S^{-1} e^{-T/S}.
\end{equation}
The typical evolution times are of the order of the number of sequences, much larger
than in the quasipecies model. This demonstrates impressively the ``guided'' nature of
quasispecies evolution \cite{eigen89}, which is much more efficient than a
random search. Figure \ref{fig_times} shows how broad the distribution (\ref{PTrecords})
is compared to that of the quasispecies model. Note, however, that for very long times
(longer than $S$) Eq.(\ref{PTrecords}) decays exponentially, faster than the
degeneracy-induced power law (\ref{power}). Taken literally, Eq.(\ref{power})
implies that the mean evolution time is infinite.

Next we consider the distribution $P_n$ of the total number $n$ of evolutionary jumps
which occur on the way to the global fitness maximum. Adapting the results of
Sibani and collaborators \cite{sibani98,sibani93}, for the case of
record dynamics we find that $P_n$ is a Poisson distribution with parameter $\ln S$,
\begin{equation}
\label{Pnrecords}
P_n \approx S^{-1} \frac{(\ln S)^{n-1}}{(n-1) !}.
\end{equation}
In Figure \ref{fig_trans} this is compared to numerical data obtained for the
quasispecies model. Again the distributions for the quasispecies
dynamics are much narrower, showing
that less transitions are required to reach the global maximum.
Simulations for longer sequences show that the mean
number of transitions increases sublinearly
in $N$, more slowly than the linear behavior 
predicted by (\ref{Pnrecords}) \cite{karl,karldip}.

\section{Outlook}
\label{Outlook}

The simplicity of the strong selection dynamics (\ref{nonlocal},\ref{local})
suggests to use it for a dynamical characterization
of different kinds of fitness landscapes. 
In contrast to the random landscapes considered here,
realistic fitness landscapes obtained e.g. from RNA folding
contain extended neutral networks in sequence space, in which the fitness
(defined in terms of the RNA secondary structure) does not change 
\cite{schuster97}. Central concepts of quasispecies theory have been
extended to such landscapes \cite{schuster01}. Extended neutrality
provides a distinct mechanism for the appearance of punctuation
patterns in evolution, since changes in the genotype do not show up in the
phenotype, as long as the former moves within a 
neutral network \cite{schuster98}.    

Another interesting direction for further research inspired by the analogy
with directed polymers is to include effects of environmental fluctuations,
which amounts to making the fitness landscape time-dependent \cite{nilsson00}.
In the directed polymer analogy, the issue is the interplay between 
{\it columnar} disorder, which is provided by the time-independent part of the landscape, and
{\it point} disorder modeling the time-dependent variations \cite{arsenin94}.
It is well known that point disorder can depin a polymer from an attractive
columnar defect in much the same way as thermal fluctuations \cite{balents94}. 
This suggests the intriguing possibility of an error threshold 
delocalization phenomenon induced by environmental fluctuations.

\vspace{0.5cm}

\noindent
{\bf Acknowledgements.} I would like to thank 
T. Halpin-Healy and C. Karl for their contributions to this project, 
and K. Sneppen, H. Flyvbjerg and L. Peliti
for useful discussions. 
This work has been supported in part by
NATO within CRG.960662, and by NSF 
under Grant No. PHY99-07949.

\end{document}